\documentclass[structabstract]{aa}  
\usepackage{graphicx}
\usepackage{txfonts}
\usepackage{natbib}
\usepackage{longtable}
\begin{document}
\title{Magnetic fields at the onset of high-mass star formation}


   \author{H.~Beuther
          \inst{1}
          \and
          J.~Soler
          \inst{1}
          \and
          W.~Vlemmings
          \inst{2}
          \and
          H.~Linz
          \inst{1}
          \and
          Th.~Henning
          \inst{1}
          \and
          R.~Kuiper
          \inst{3}
          \and
          R.~Rao
          \inst{4}
          \and
          R.~Smith
          \inst{5}
          \and
          T.~Sakai
          \inst{6}
          \and
          K.~Johnston
          \inst{7}
          \and
          A.~Walsh
          \inst{8}
          \and
          S.~Feng
          \inst{9}
}
   \institute{$^1$ Max Planck Institute for Astronomy, K\"onigstuhl 17,
              69117 Heidelberg, Germany, \email{beuther@mpia.de}\\
              $^2$ Department of Earth and Space Sciences Chalmers University of Technology, Onsala Space Observatory, 439 92, Onsala, Sweden\\
              $^3$ Institute of Astronomy and Astrophysics, University of T\"ubingen, Auf der Morgenstelle 10, 72076, T\"ubingen, Germany\\
              $^4$ Academia Sinica Institute of Astronomy and Astrophysics, 645 N. A’ohoku Place, Hilo, HI 96720, USA 0000-0002-1407-7944\\
              $^5$ School of Physics and Astronomy, University of Manchester, Oxford Road, Manchester M13 9PL, UK\\
              $^6$ Department of Communication Engineering and Informatics, The University of Electro-Communications, Chofugaoka, Chofu, Tokyo 182-8585, Japan\\
              $^7$ School of Physics and Astronomy, University of Leeds, Leeds, LS2 9JT, UK\\
              $^8$ International Centre for Radio Astronomy Research, Curtin University, GPO Box U1987, Perth WA 6845, Australia\\
              $^9$ Max-Planck-Institut f\"ur Extraterrestrische Physik, Giessenbachstrasse 1, 85748 Garching, Germany
}

   \date{Version of \today}

\abstract
{The importance of magnetic fields at the onset of star formation
  related to the early fragmentation and collapse processes is largely
  unexplored today.}
{We want to understand the magnetic field properties at the earliest
  evolutionary stages of high-mass star formation.}
{The Atacama Large Millimeter Array is used at 1.3\,mm wavelength in
  full polarization mode to study the polarized emission and by that the
  magnetic field morphologies and strengths of the high-mass starless
  region IRDC\,18310-4.}
{The polarized emission is clearly detected in four sub-cores of the
  region. In general it shows a smooth distribution, also along
  elongated cores. Estimating the magnetic field strength via the
  Davis-Chandrasekhar-Fermi method and following a structure function
  analysis, we find comparably large magnetic field strengths between
  $\sim$0.6 and 3.7\,mG. Comparing the data to spectral line
  observations, the turbulent-to-magnetic energy ratio is low,
  indicating that turbulence does not significantly contribute to the
  stability of the gas clump. A mass-to-flux ratio around the critical
  value 1.0 -- depending on column density -- indicates that the region
  starts to collapse which is consistent with the previous spectral
  line analysis of the region.}
{While this high-mass region is collapsing and thus at the verge of
  star formation, the high magnetic field values and the smooth
  spatial structure indicate that the magnetic field is important for
  the fragmentation and collapse process. This single case study can
  only be the starting point for larger sample studies of magnetic
  fields at the onset of star formation.}  

\titlerunning{Magnetic fields at the onset of high-mass star formation}

\maketitle

\section{Introduction}
\label{intro}

The importance of magnetic fields during the formation of stars has
been topic of great controversy over the last decades. While some
groups stress that magnetic fields have to be important during cloud
formation and core collapse processes (e.g.,
\citealt{mouschovias1979,mouschovias2006,commercon2011,peters2011,tan2013,tassis2014,myers2013,myers2014}),
others consider that the effects of turbulence and gravity are far
more important for governing star formation (e.g.,
\citealt{padoan2002,klessen2005,vazquez2011}). Even the interpretation
of a single dataset can be extremely controversial regarding the
importance of magnetic fields (e.g.,
\citealt{mouschovias2009,crutcher2010}).

More specifically, at the onset of collapse during the formation of
high-mass clusters, observations show that gas clumps fragment,
however, on average less than predicted by classical Jeans
fragmentation (e.g.,
\citealt{wang2014,beuther2015a,zhang2015,fontani2016}).  Different
processes are able to explain the suppressed fragmentation of the
initial gas clumps, in particular steep initial density structures
(e.g., \citealt{girichidis2011}), turbulence (e.g.,
\citealt{wang2014}) or magnetic fields (e.g.,
\citealt{commercon2011,fontani2016,klassen2017}).

\begin{figure*}[htb]
\includegraphics[width=1.0\textwidth]{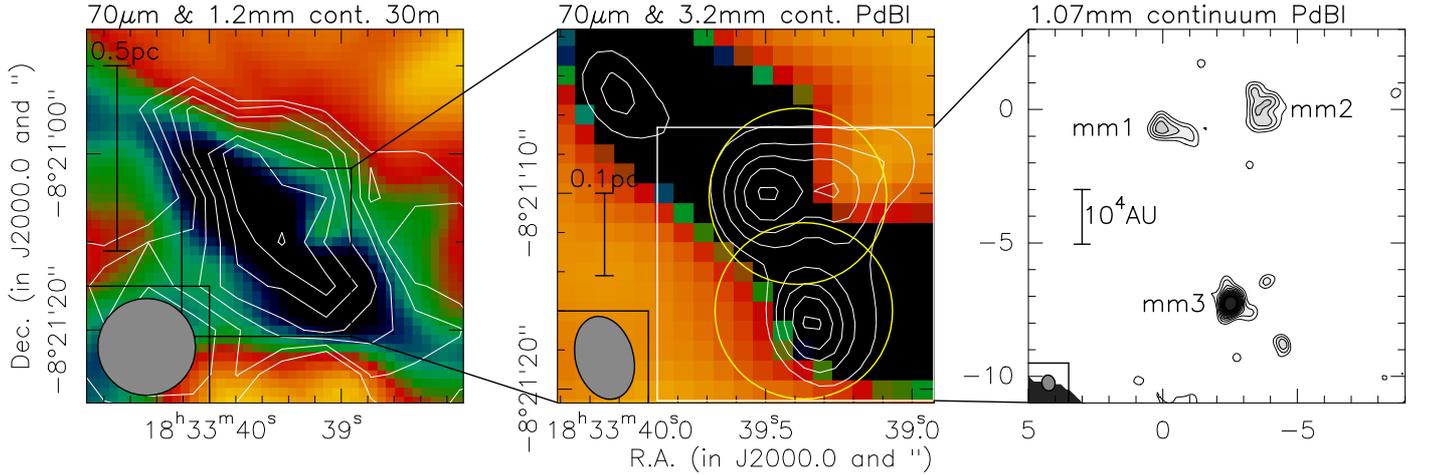}
\caption{Compilation of the continuum images in IRDC\,18310-4
  \citep{beuther2013a,beuther2015a}. The left and middle panels show
  in color the Herschel 70\,$\mu$m image with a stretch going dark for
  low values. The white contours in the left panel show the 1.2\,mm
  MAMBO continuum observations starting from 4$\sigma$ and continuing
  in 1$\sigma$ steps with a $1\sigma$ value of 13\,mJy\,beam$^{-1}$.
  The contours in the middle panel presents the 3.2\,mm PdBI continuum data
  starting from 3$\sigma$ and continuing in 2$\sigma$ steps with a
  $1\sigma$ value of 0.08\,mJy\,beam$^{-1}$. The two yellow circles
  outline the $\sim$9$''$ inner third of the two pointings (field 1
  north, field 2 south) we observed in this polarization project. The
  right panel then shows the new 1.07\,mm ALMA continuum observations
  starting from 3$\sigma$ and continuing in 1$\sigma$ steps with a
  $1\sigma$ value of 0.6\,mJy\,beam$^{-1}$.}
\label{18310}
\end{figure*}

The Atacama Large Millimeter Array (ALMA) with its polarization
capabilities now allows us to address the magnetic field aspect
observationally in depth. To investigate the magnetic field properties
at the onset of high-mass star formation, we selected the infrared
dark cloud IRDC18310-4 located at an approximate distance of 4.9\,kpc
\citep{sridharan2005}. The region is infrared dark even at 70\,$\mu$m
wavelengths, has a mass reservoir of $\sim$800\,M$_{\odot}$ within
roughly a square parsec, and shows no signs of active star formation
(Fig.~\ref{18310} shows a compilation of previous data). The region
was studied with the Plateau de Bure Interferometer in the 3\,mm and
1\,mm band, and hierarchical fragmentation on increasingly smaller
scales down to $\sim 2500$\,AU was identified
\citep{beuther2015a}. Spectral line signatures indicate that the whole
maternal gas is likely globally collapsing, fragmenting and at the
onset of star formation \citep{beuther2013a,beuther2015a}. The
fragment separations in that region are consistent with thermal Jeans
fragmentation, however, the core masses are more than an order of
magnitude larger than the typical Jeans mass
\citep{beuther2015a}. They suggest that this discrepancies can be
reconciled by invoking either non-homogeneous initial density
structures or strong magnetic fields. Here, we are investigating the
magnetic field properties of the region.

\section{Observations}
\label{obs}

The infrared dark cloud IRDC18310-4 was observed with the Atacama
Large Millimeter Array (ALMA) as a cycle 3 project (2015.1.00492.S)
during four sessions between June 17 and June 20, 2016. The
observations were conducted in the 1\,mm band with the at that time
for ALMA polarization observations still predefined frequency setting
with the LO at 233\,GHz to avoid any strong line contamination. The
four basebands with 2\,GHz bandwidth each were centered at 224, 226,
240 and 242\,GHz. The channel width was 31.25\,MHz. At this frequency,
the primary beam of the ALMA 12\,m array is $27''$, in principle large
enough to encompass our region of interest. However, taking into
account that ALMA still recommends that only the inner third of the
primary beam is reliable for polarization measurements, we used two
fields centered on the main continuum sources as shown in
Fig.~\ref{18310}. The two corresponding phase centers are for field 1
R.A. (J2000.0) 18:33:39.4, Dec. (J2000.0) -08:21:10.2 and for field 2
R.A. (J2000.0) 18:33:39.4, Dec. (J2000.0) -08:21:16.0, separated by
only $5.8''$ in declination.  Two slightly different array
configurations were used (C36-2, C36-3) with a total baseline range
between 15 and 704\,m. The shortest baseline would correspond
theoretically to maximum recoverable scales of approximately
$22''$. However, typically such theoretical limits are not achieved in
interferometric imaging because of the weighting of data and missing
baselines also at intermediate scales. ALMA reports for the given
configurations maximum recoverable scales of $\sim$11$''$ at
230\,GHz\footnote{https://almascience.eso.org/observing/prior-cycle-observing-and-configuration-schedule}. Hence,
large-scale flux is still filtered out which will be quantified in the
coming section. Eight executions (two per session) of the same
scheduling block were conducted with on average 35 good antennas in
the array. The execution times per scheduling block varied between 85
and 99\,min, and the on-source integration time for each observed
field was $\sim$17.6\,min per execution. Bandpass calibration was
conducted with observations of the quasar J1751+0939 whereas the
absolute flux was calibrated with J1733-1304 or Titan. The absolute
flux uncertainty should be around 10\%. To calibrate phases and
amplitudes, regularly interleaved observations of the quasar
J1912-0804 were used. The polarization calibration was conducted with
J1743-0350.

\begin{figure*}[htb]
\includegraphics[width=1.0\textwidth]{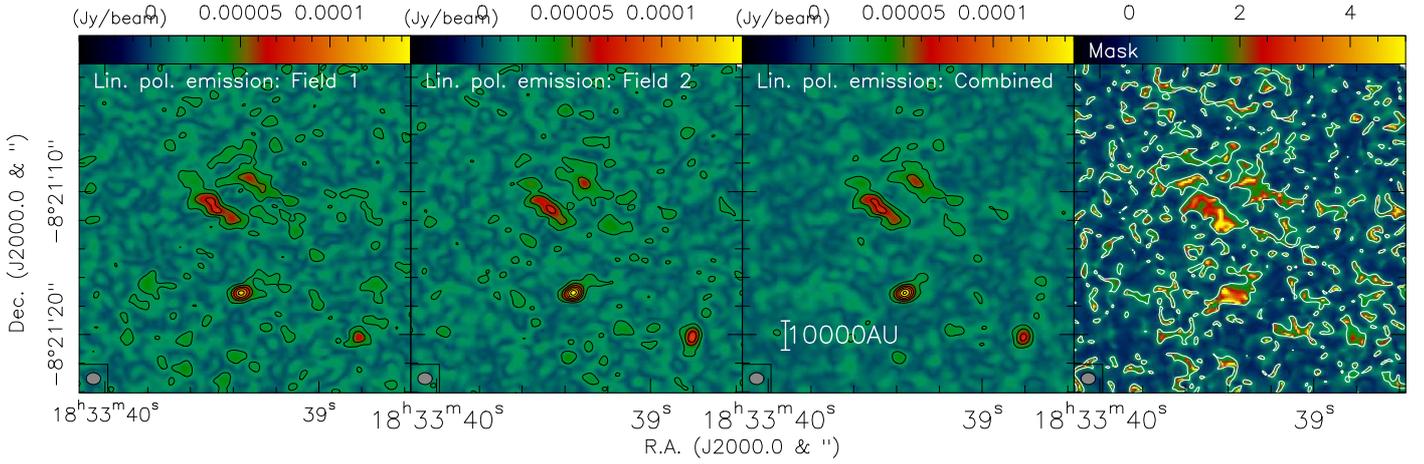}
\caption{Comparison of the individually and combined two fields. The
  left panel shows the linearly polarized emission using only field 1,
  the second panel presents the same for using only field 2, and the
  third panel shows the combined image. The contour levels are always
  chosen as the 4$\sigma$ levels of the combined image where 1$\sigma$
  is 6\,$\mu$Jy\,beam$^{-1}$. The right panel presents the mask
  created from the combined image and the residuals of each individual
  image following \citet{matthews2014}, see section \ref{obs}. The
  contour level is set to 1, corresponding to the $3\sigma$ confidence
  level in \citet{matthews2014}. The beam sizes are shown at the
  bottom-left of each panel, and the combined image also shows a scale
  bar.}
\label{compare}
\end{figure*}

Calibration and imaging of the data was done in CASA version 4.7. For
the calibration we followed the provided procedures from the ALMA
observatory. Because the two target fields are so closely spaced, we
want to combine them in a single image. However, since the main
sub-structures are partly outside the recommended inner third of the
primary beam (Fig.~\ref{18310}), tests of the individually imaged
fields and combined images were conducted. Figure \ref{compare} shows
the linearly polarized images of the region for each of the two fields
separately as well as for the combined dataset. The structures between
the two individually imaged fields agree well, and the combined image
has the expected lower rms. We conducted an additional test following
the approach outlined in \citet{matthews2014} where the residuals of
the linearly polarized Q and U components ($Q_{\rm{res}}$ \& and
$U_{\rm{res}}$) are calculated first. Then the combined image
($\sqrt{Q^2+U^2}$) is compared to the residual image to gauge the
trustworthiness of the combination. In practice we calculated
\begin{eqnarray}
Q_{\rm{res}}=(Q_1-Q_2)/2 \hspace{0.3cm} \& \hspace{0.3cm} U_{\rm{res}}=(U_1-U_2)/2 \\
Mask = \frac{\sqrt{Q^2+U^2}}{3\times \sqrt{Q^2_{\rm{res}}+U^2_{\rm{res}}}}
\end{eqnarray}
In equation 2, the mask is created for the combined linear polarized
image, which is above three times the combined residual image. The
corresponding mask-image is shown in the right panel of
Fig.~\ref{compare}. Clearly, the main structures in the polarized
intensity map are several times above the here calculated $3\sigma$
value, showing again that the combination of the two fields is a valid
approach.

Based on these tests, we imaged both datasets together in a mosaic mode
with natural weighting and a small uv-taper on the longer baselines to
increase our signal-to-noise ratio. All four spectral windows were
used for the imaging process. The synthesized beam of these images is
$1.01''\times 0.83''$ (PA 89$^o$). The $1\sigma$ rms values of the
Stokes I and the linear polarized image are 0.15\,mJy\,beam$^{-1}$ and
6\,$\mu$Jy\,beam$^{-1}$, respectively.

\section{Results}
\label{results}

\begin{figure*}[htb]
\includegraphics[width=1.0\textwidth]{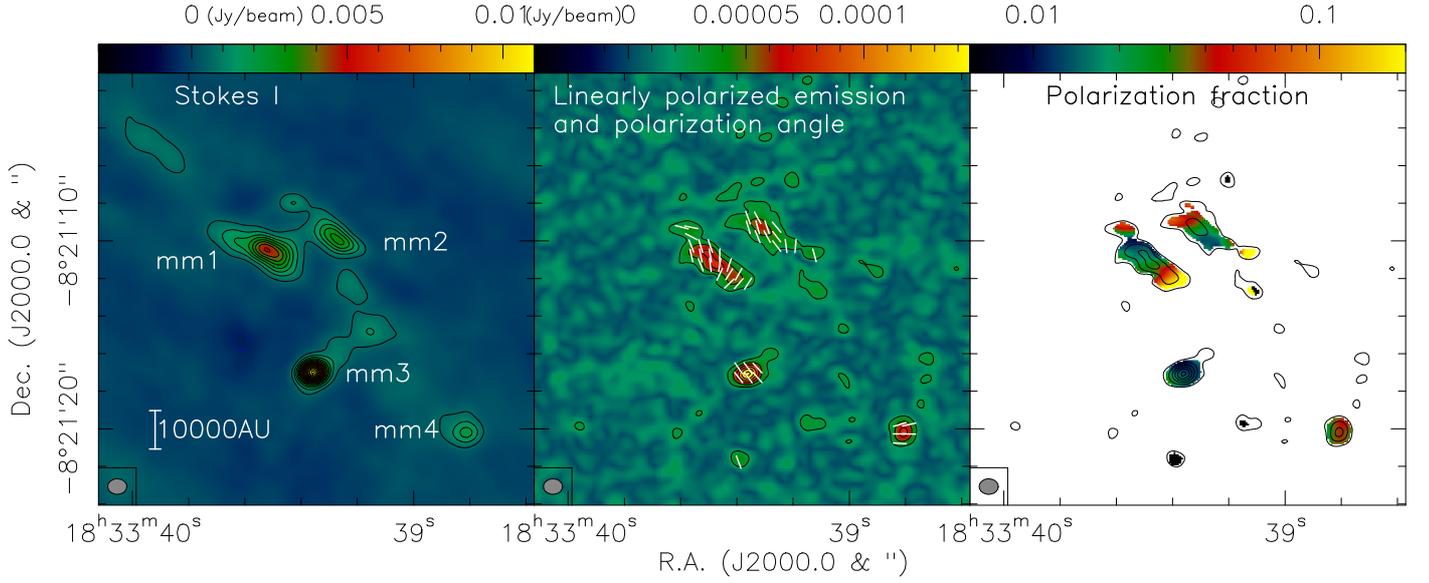}
\caption{Polarization data at 1.3\,mm wavelength for IRDC18310-4. The
  left panel shows the Stokes I emission ($4\sigma$ contours of
  0.6\,mJy\,beam$^{-1}$), and the middle panel presents the linearly
  polarized emission ($4\sigma$ contours of
  24\,$\mu$Jy\,beam$^{-1}$). The line segments show the corresponding
  polarization angles (the magnetic field should be rotated by
  90\,$^{\circ}$). The right panel shows for comparison the polarization
  fraction in color with the same contours of linearly polarized
  emission as in the middle panel. The polarization fraction should be
  considered as upper limit because Stokes I may be more strongly
  affected by interferometric spatial filtering than the weaker
  polarized emission.}
\label{overview}
\end{figure*}

The following analysis is based on the combined dataset of both
observed fields. Fig.~\ref{overview} presents an overview of the
obtained data. The left panel shows the Stokes I image of the region,
and we clearly re-identify the three main continuum peaks known from
the previous PdBI observations (Fig.~\ref{18310}). In addition to
these main sources, the more sensitive ALMA data identify additional
dust continuum sources in the field. Particularly strong is the
south-western source mm4, whereas some weaker features are found
between mm2 and mm3 as well as at the north-eastern edge of the
field. We will concentrate the study of the linearly polarized
emission on the four strongest sources mm1 to mm4. The peak
($S_{\rm{peak}}$) and integrated fluxes ($S_{\rm{int}}$) within the
$4\sigma$ contours are shown in Table \ref{para}. The total integrated
flux density of the whole region is recovered in the Stokes I image of
Fig.~\ref{overview} is 72\,mJy\,beam$^{-1}$. For comparison the
single-dish 1.2\,mm bolometer flux density measured with the MAMBO
camera at the IRAM 30\,m telescope \citep{beuther2002a} within the
inner $11''$ beam size is 132\,mJy, and the 1.2\,mm single-dish MAMBO
flux density measured within the ALMA primary beam area is
$\sim$410\,mJy. Taking into account that there is a slight difference
in the observed wavelength between the 1.2\,mm single-dish and the
1.3\,mm of the ALMA data approximately 20-50\% of the total flux are
recovered by our interferometer observations. Assuming optically thin
dust emission and following the approach already used in
\citet{beuther2013a,beuther2015a} we can calculate the H$_2$ column
densities (N$_{\rm{H}_2}$) and gas masses for the four
sub-regions. With gas temperatures in this starless core around
$\sim$15\,K, a gas-to-dust mass ratio of 150 \citep{draine2011}, and a
dust opacity index $\kappa$ discussed in \citet{ossenkopf1994} for
grains with thin ice mantles at densities of $10^6$\,cm$^{-3}$
($\kappa_{1.3\rm{mm}}\sim 0.9$\,cm$^2$g$^{-1}$), the derived column
densities and masses are presented in Table \ref{para}. With the main
uncertainties in the dust model and the temperature, we estimate the
accuracy of the masses and column densities within a factor 2. The
core masses range between roughly 4 and 22\,M$_{\odot}$ and the H$_2$
column densities between $2.4\times 10^{23}$ and $1.36\times
10^{24}$\,cm$^{-2}$, consistent with the previous findings in
\citet{beuther2015a}.

\begin{table}[htb]
\caption{Source parameters}
\begin{tabular}{lrrrr}
\hline \hline
 & $S_{\rm{peak}}$ & $S_{\rm{int}}$ & N$_{\rm{H}_2}$ & $M$ \\
 & $\left(\frac{\rm{mJy}}{\rm{beam}}\right)$ & (mJy) & ($10^{23}$cm$^{-2}$) & (M$_{\odot}$) \\
\hline
mm1 & 4.9 & 19.4 & 5.7 & 22.4 \\
mm2 & 3.2 & 8.3  & 3.8 & 9.6 \\
mm3 & 1.2 & 14.4 & 13.6& 16.7 \\
mm4 & 2.1 & 3.9  & 2.4 & 4.5 \\
\hline \hline
\end{tabular}
\label{para}
\end{table}

In the context of the magnetic field investigation even more
important, Fig.~\ref{overview} presents in the middle panel the
linearly polarized emission from the region. All four main mm
continuum sources are clearly detected in the polarized
emission. Fig.~\ref{overview} also presents the observed polarization
angles, and within each of the four cores, the polarization angles
exhibit a relatively ordered structure. While for the more compact
sources mm3 and mm4 less independent polarization angle measurements
are possible, the orientation in the two sources is roughly in
northeast-southwest and east-west directions for the two regions,
respectively. For the more extended sources mm1 and mm2, the
polarization angle distribution shows some smooth angle shifts over
the extents of the structures. The well-structured polarization angles
are already a first indication that the magnetic field may be
important in that region because turbulence-dominated regions would
show less structured polarization angle distributions. We also do not
see signs of polarization holes toward the peak positions like those
observed previously toward low-mass star-forming regions (e.g.,
\citealt{wolf2003c,hull2014}). However, that fact that the mean
polarization angles toward the four main mm sources are very different
may be one explanation for the sometimes observed polarization holes
in single-dish data (e.g., \citealt{matthews2009}). Single-dish
observations would average over the polarization properties of the
four sub-sources which could result in a lowered signal toward the
lower-resolution single-dish data. In addition to this, the right
panel in Fig.~\ref{overview} presents the corresponding polarization
fraction, and we find values typically between 1 and 10\%. The latter
high values should be considered as upper limits because the much
stronger Stokes I emission, that may come from comparably larger
scales, may be more strongly affected by interferometric spatial
filtering than the weaker polarized emission. While we can measure the
missing flux for Stokes I (see above), we do not have the single-dish
information for the linearly polarized emission. Hence, we cannot
conclusively answer how much missing flux differences between
polarized and non-polarized emission may affect the polarization
fraction measurement. For comparison, \citet{hull2014} measured in
their TADPOL magnetic field study of 30 mainly low-mass star-forming
regions (with a few high-mass more evolved regions included) typical
polarization fractions between 1 and $\sim$8\%, largely consistent
with our measurement. However, they also find at $2.5''$ regularly
polarization holes toward the peak positions which is not
evident in our data. This indicates that at the scales resolved here
for IRDC\,18310-4, the magnetic field structure is still largely
ordered and coherent.

\section{Magnetic fields}

\subsection{Magnetic field strength via  Davis-Chandrasekhar-Fermi method}

The Davis-Chandrasekhar-Fermi method \citep[hereafter the
DCF,][]{davis1951,chandrasekhar1953} can be used to calculate the
magnetic field strength in a gas if the angular dispersion of the
local magnetic field orientations $\sigma_{\psi}$, the gas density
$\rho$, and the one dimensional velocity dispersion of the gas
$\sigma_{v}$, are known.  Assuming that the magnetic field is frozen
into the gas and that the dispersion of the local magnetic field
orientation angles is due to transverse and incompressible Alfv\'{e}n
waves, then the strength of the plane-of-the-sky component of the
magnetic field is

\begin{equation}
B^{\rm{DCF}}_{\perp}=\sqrt{4\pi\rho}\frac{\sigma_{v}}{\sigma_{\psi}}.
\end{equation}

We calculate $B^{\rm{DCF}}_{\perp}$ using the procedure described in
appendix D of \cite{planckXXXV}.  We estimate $\rho$ assuming an
average number density $n=10^6$\,cm$^{-3}$ \citep{beuther2013a} and a
mean molecular weight $\mu=2.8\,m_{p}$, where $m_{p}$ is the proton
mass. The one-dimensional velocity dispersion $\sigma_{v}$ is
estimated from the width of the N$_2$H$^+$ line measured in this
region \citep{beuther2015a}. With a line width $\Delta
v(\rm{N_2H^+})\sim 0.3$\,km\,s$^{-1}$, we get $\sigma_{v}\approx
\Delta v(\rm{N_2H^+})/\sqrt{8\rm{ln}2}\approx
0.3\rm{km/s}/\sqrt{8\rm{ln}2}\approx 0.13$\,km\,s$^{-1}$. Because more accurate, individual estimates per core are difficult, we use the same value of $\sigma_{v}$ and $n$ for all three sub-regions. We estimate the values of the
angular dispersion of the local magnetic field orientations directly
from the Stokes $Q$ and $U$ by using

\begin{equation}\label{eq:SigmaPsi}
\sigma_{\psi}=\sqrt{\left<(\Delta\psi)^{2}\right>},
\end{equation}

where

\begin{equation}
\Delta\psi=0.5\times\arctan\left(\frac{Q\left<U\right>-\left<Q\right>U}{Q\left<Q\right>-\left<U\right>U}\right)
\end{equation}

and $\left<\cdots\right>$ denotes an average over the selected pixels
in each map. The angular dispersions of the polarization vectors as
well as the derived magnetic field values are presented in Table
\ref{table:DCFvalues} for mm1 to mm3 (mm4 is too close to a point
source for reasonable estimates). On average we find comparably high
magnetic field values in the several hundred $\mu$G to mG regime.

The formal errors are calculated from the $4\sigma$ rms values from
the individual Stokes Q and U maps and the area of the
calculations. The error for mm3 is largest because it is the smallest
source and hence dominated by beam effects. One should keep in mind
that the absolute errors for the Davis-Chandrasekhar-Fermi method are
even larger because of its underlying assumptions and the effect of
projection and line-of-sight integration, which are discussed in
detail in appendix D of \citet{soler2016}. For example, it is
difficult to determine whether the measured dispersion around the mean
field $\sigma_{\psi}$ is exclusively the effect of
magneto-hydrodynamic waves and turbulence. Moreover, the observed
value of $\sigma_{\psi}$ is an average of various magnetic field
vectors, and integration of multiple components along the line may
decrease the measurable polarization angle compared to the true
dispersion of the magnetic field orientation.  Furthermore, the
velocity dispersion estimates are obtained using a tracer that does
not necessarily sample the same volume as traced by the dust
polarization. Finally, the mean densities are assumed whereas density
gradients exist within each region.  In summary, the magnetic field
estimates from the Davis-Chandrasekhar-Fermi method should be
considered as order-of-magnitude estimates, which are illustrative of
the differences between the regions, but do not fully encompass the
complexity of the field dynamics in each one of them.

\begin{table}
\caption{Magnetic field strength estimates}              
\label{table:DCFvalues}      
\centering                                      
\begin{tabular}{lcccc}          
\hline\hline                        
Source & $\sigma_\psi$ & $B^{\rm{DCF}}_{\perp}$ & $b(0)$ & $B^{\sc SF}_{\perp}$ \\   
       & [deg]        & [mG]               & [deg]  & [mG] \\
\hline                                   
mm1 &23.1 & 0.3$\pm$0.1 & 1.6 & 3.7$\pm$1.3\\ 
mm2 & 9.8 & 0.6$\pm$0.2 & 8.8 & 0.6$\pm$0.2\\
mm3 & 4.3 & 1.3$\pm$0.4 & 3.3 & 1.7$\pm$0.6\\
\hline                                             
\end{tabular}
\end{table}

\subsection{Structure function of the polarization orientation angles}

\begin{figure}[htb]
\includegraphics[width=0.49\textwidth]{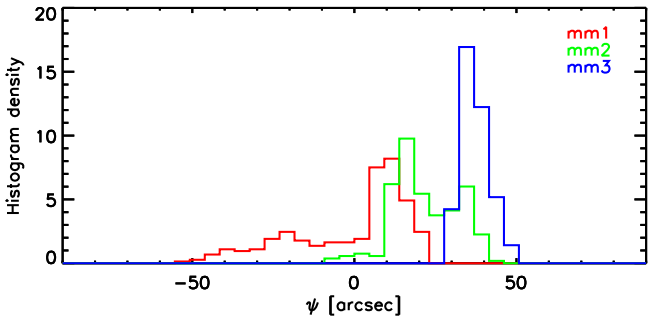}
\caption{Histogram of polarization orientation angles, $\psi$, towards
  each of the sources.}
\label{fig:histPsi}
\end{figure}

In dense clouds, the magnetic field structure is the result of
multiple physical processes not accounted for by the basic DCF
analysis.  Consequently, dispersion measured about mean fields,
assumed straight, may be much larger than should be attributed to
magneto-hydrodynamical waves or turbulence.  In the region
IRDC\,18310-4, the distribution of polarization orientation angles,
illustrated in Fig.~\ref{fig:histPsi}, shows that in the sources mm1
and mm2 there is broad distribution or two separate components of
polarization orientations, which could lead to an overestimation of
$\sigma_\psi$ and, correspondingly, an underestimation of $B^{\sc
  DCF}_{\perp}$.

Following \citet{hildebrand2009} and \citet{houde2009,houde2011}, we
can analyze the structure function (of second order) $S_2(\ell)$, also
known as the dispersion function, of the polarization angles. This
structure function essentially measures the differences between
polarization angles separated by displacements $\ell$. Large values in
$S_2$ reflect large variations whereas small values indicate less
dispersion between measured polarization angles. This structure
function of the polarization orientation angles $S_{2}(\ell)$ allows
for the evaluation of the dispersion of polarization angles,
$\sigma_{\psi}$, while avoiding the effect of large-scale
non-turbulent perturbations \citep{hildebrand2009}.  The evaluation of
$\sigma_{\psi}$ is made by fitting the square of the structure
function of polarization angles, $S_{2}(\ell)$, with a second-order
polynomial, $S^{2}_{2}(\ell)=b(\ell)+a'_{2}\ell^{2}$ above length
scales that are dominated by the beam, in this case we chose to fit
for $\ell>0.5''$. Evaluating the intercept $b(0)$ of that fit gives an
alternative measure for the dispersion of polarization angles.

\begin{figure}[htb]
\includegraphics[width=0.49\textwidth]{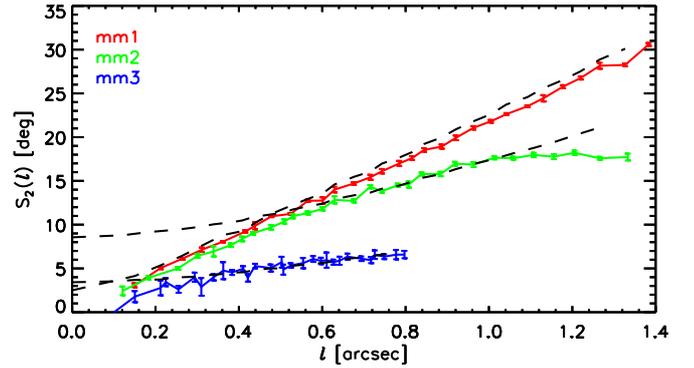}
\caption{Structure function of the polarization orientation angles,
  $S_{2}(\ell)$, towards each of the sources.  The dashed lines
  correspond to the fit to the values of $S_{2}(\ell)$ in the range
  $\ell>0.5''$.} 
\label{structure_function}
\end{figure}

It is evident from the values of $S_{2}(\ell)$, presented in
Fig.~\ref{structure_function}, that the three sources have very
different behaviors.  In the case of source {\it mm3}, the available
data does not allow to extend the study of $S_{2}(\ell)$ to
$\ell$-values larger than the size of synthesized beam, thus it seems
to be dominated by the interferometer resolution.  In the case of
source mm2, the values of $S_{2}(\ell)$ lead to a value of
$b(0)$ that is very close to the values of $\sigma_{\psi}$ estimated
using Eq.~\ref{eq:SigmaPsi}, suggesting that despite the two peaks
in the orientation angle distribution the assumption of a single
straight mean magnetic field direction is not too inadequate for this
region.

The case of source mm1 is, however, very different from the that of
mm2 and mm3.  There, the relatively large values of $\sigma_{\psi}$
reflects the broad orientation angle distribution, but in contrast to
mm2, the values $S_{2}(\ell)$ of the dispersion of polarization
orientation angles that can be directly attributed to turbulence is
very small, that is, $b(0)=$1.6\,deg.  This indicates that in mm1 the
mean magnetic field direction is not straight and that despite the
conclusions that can be drawn from the simple DCF, the magnetic field
has to be comparatively stronger. The values of the magnetic field
5strengths based on the angle dispersions calculated from
$S_{2}(\ell)$, $B^{\sc SF}_{\perp}$ are also summarized in
Table~\ref{table:DCFvalues}.
 
\subsection{Magnetic field implications}

With an approximate mean magnetic field strength in the plane of the
sky of 2.0\,mG ($B^{\sc SF}_{\perp}$ values in Table
\ref{table:DCFvalues}) and a typical density of $10^6$\,cm$^{-3}$
\citep{beuther2013a}, the one-dimensional Alfven velocity is $\sigma_A
= B/\sqrt{4\pi\rho}\sim 2.6$\,km\,s$^{-1}$. In the following we assume
that the one-dimensional Alfven velocity in the plane of the sky is
the same as that along the line of sight. For comparison, the
one-dimensional velocity dispersion for individual cores is
$\sigma_{v}\sim$0.13\,km\,s$^{-1}$ (see above,
\citealt{beuther2015a}). Or using the approximate line width of the
whole maternal clump of $\Delta v\sim 1.5$\,km\,s$^{-1}$
\citep{beuther2013a}, the corresponding one-dimensional velocity
dispersion is $\sim$0.64\,km\,s$^{-1}$. These numbers indicate that
any turbulent or infall velocity is below the Alfvenic
velocity. Following \citet{girart2009} or \citet{beuther2010c}, we can
estimate the ratio of turbulent-to-magnetic energy as $\beta \sim
3(\sigma_{v}/\sigma_A)^2$. Using the larger line-of-sight velocity
dispersion of the whole clump that results in turbulent-to-magnetic
energy ratio of $\beta \approx 0.18$ whereas the lower $\sigma_{v}$
for the individual cores would result in $\beta \approx 0.008$. Hence,
for this high-mass starless region, the magnetic energy appears to
dominate over the turbulent energy and thus should be an important
ingredient for the formation of high-mass stars.

In our previous study of the region \citep{beuther2015a}, we discussed
whether thermal Jeans fragmentation can explain the general
fragmentation properties of the region or whether an adapted turbulent
Jeans fragmentation may be more likely. While the fragment separations
were found to be consistent with thermal Jeans fragmentation, the
fragment masses are much higher than the thermal predictions
\citep{beuther2015a}. Although turbulent Jeans fragmentation could
explain the higher masses, the fact that the measured line width
toward individual cores are so small \citep{beuther2015a} does make
the turbulent interpretation less likely. In that context, it is
interesting to investigate how the magnetic field would affect such
estimates. If one replaces the thermal sound speed at the given low
temperatures of $\sim$15\,K by the Alfven velocity estimated above,
the estimated length scales would rise by up to a factor 20, and the
estimated masses by even the cube of that. Such high separations and
masses exceed the observed values \citep{beuther2015a}. Hence, while
the magnetic field definitely influences the fragmentation of the
region, such a simplified ``magnetic Jeans fragmentation'' picture
cannot explain the observed data alone.

A different way to gauge the stability of a region is the mass-to-flux
ratio $M/\Phi_B\approx 7.6\times 10^{-24}\frac{N_{\rm{H_2}}}{B}$ that
is given in units of the critical mass-to-flux ratio
$(M/\Phi_B)_{\rm{crit}}$ (with $N_{\rm{H_2}}$ and $B$ in cm$^{-2}$ and
mG; \citealt{crutcher1999,troland2008}). Using the average magnetic
field of $\sim$2.0\,mG and the column density $N_{\rm{H_2}}\sim
1.3\times 10^{23}$\,cm$^{-2}$ observed on larger-scales ($11''$) with
single-dish instruments \citep{beuther2013a}, we get a mass-to-flux
ratio $M/\Phi_B$ of $\sim$0.5. This is likely a lower limit
because it may well be that the magnetic field strength on larger
spatial scales could be lower than what we measure at the higher inner
densities. Using the higher average column densities from these ALMA
observations $N_{\rm{H_2,av}} \sim 8.2\times 10^{23}$\,cm$^{-2}$, one
gets $M/\Phi_B\sim 3.1$. This indicates that while on large scales the
overall gas clump may still be at the verge of criticality, on smaller
scales it should be collapsing. Such early collapse motions are also
inferred from the spectral line data in \citet{beuther2015a}.

\section{Discussion and conclusions}
\label{conclusion}

To the authors knowledge, this is the first high-spatial-resolution
magnetic field study of a very young high-mass star-forming region at
the onset of collapse prior to any signpost of star formation. For
more evolved regions hosting high-mass protostellar objects and/or hot
molecular cores, the findings about the importance of the magnetic
field are ambiguous. While for some regions the magnetic field should
play a dominant role (e.g., G31.4, W75N, W51 or CepA;
\citealt{girart2009,surcis2009,tang2009a,vlemmings2010}), other
systems appear to be more strongly influenced by dynamic motions
(e.g., IRAS\,1809-1732; \citealt{beuther2010c}). The magnetic field
values estimated for IRDC\,18310-4 on the order of $\sim$2\,mG are
lower than what is found for several more evolved hot-core like
regions, e.g., W3(H$_2$O), IRAS\,18089-1732 or NGC7538IRS1 with
reported values of 17, 11 and 2.5\,mG, respectively
\citep{chen2012,beuther2010c,frau2014}. With only one high-mass
starless region observed so far, this difference is not conclusive
yet, but it will be interesting to see with future data whether the
measured magnetic field strengths are indeed related to the
evolutionary stage.

Our finding for this high-mass starless region that the mass-to-flux
ratio is of order unity (depending on the column density) and at the
same time the turbulent-to-magnetic energy ratio is comparably low are
both indications for the general instability of the region. The latter
fact mainly implies that turbulence is not contributing significantly
to the stability of the region.  But stability is not needed anyway,
because the kinematic signatures found by \citet{beuther2015a} already
implied that the whole gas clump is globally collapsing and at the
verge of star formation. This is consistent with the high mass-to-flux
ratio. The mean magnetic field strength around $\sim$2.0\,mG does not
contradict that picture, it just implies that the magnetic field
significantly contributes to the fragmentation and collapse properties
of the overall gas clump. For example, high magnetic field values can
inhibit fragmentation into many low-mass cores (e.g.,
\citealt{commercon2011,peters2011,myers2013,myers2014,fontani2016}). For
IRDC\,18310-4, even at the higher resolution of the previous PdBI
observations, several of the cores do not fragment but show masses in
excess of 10\,M$_{\odot}$ \citep{beuther2015a}. Although the
sensitivity of these data is not sufficient to detect cores below
1\,M$_{\odot}$, the fact that massive non-fragmenting cores at the
given spatial resolution are identified, is indicative of large
magnetic fields that we observe now with the new ALMA data. The
overall smooth spatial structure of the polarization angles is
additional evidence for the dynamic importance of the magnetic field.

In summary, while the maternal gas clump is collapsing at large, the
polarization data reveal that the magnetic field is important for the
fragmentation and star formation process in this region. However, a
single case study is not sufficient for a general conclusion, and it
can only set the stage for future studies of larger samples.

\begin{acknowledgements} 
  This paper makes use of the following ALMA data:
  ADS/JAO.ALMA\#2015.1.00492.S. ALMA is a partnership of ESO
  (representing its member states), NSF (USA) and NINS (Japan),
  together with NRC (Canada) and NSC and ASIAA (Taiwan) and KASI
  (Republic of Korea), in cooperation with the Republic of Chile. The
  Joint ALMA Observatory is operated by ESO, AUI/NRAO and NAOJ. HB,
  AA, JM and FB acknowledge support from the European Research Council
  under the Horizon 2020 Framework Program via the ERC Consolidator
  Grant CSF-648505. WV acknowledges funding from the European Research
  Council under the European Union’s Seventh Framework Programme
  (FP/2007-2013) / ERC Grant Agreement n. 614264.  RK acknowledges
  financial support from the German Research Foundation (DFG) under
  the Emmy Noether Research Program via the grant no.~KU 2849/3-1. RS
  acknowledges support from the STFC through an Ernest Rutherford
  Fellowship.
\end{acknowledgements}

\end{document}